# Jointly Optimal Channel Pairing and Power Allocation for Multichannel Multihop Relaying

Mahdi Hajiaghayi, *Student Member, IEEE*, Min Dong, *Senior Member, IEEE*, and Ben Liang, *Senior Member, IEEE*

*Abstract*—We study the problem of channel pairing and power allocation in a multichannel multihop relay network to enhance the end-to-end data rate. Both amplify-and-forward (AF) and decode-and-forward (DF) relaying strategies are considered. Given fixed power allocation to the channels, we show that channel pairing over multiple hops can be decomposed into independent pairing problems at each relay, and a sorted-SNR channel pairing strategy is sum-rate optimal, where each relay pairs its incoming and outgoing channels by their SNR order. For the joint optimization of channel pairing and power allocation under both total and individual power constraints, we show that the problem can be decoupled into two subproblems solved separately. This separation principle is established by observing the equivalence between sorting SNRs and sorting channel gains in the jointly optimal solution. It significantly reduces the computational complexity in finding the jointly optimal solution. It follows that the channel pairing problem in joint optimization can be again decomposed into independent pairing problems at each relay based on sorted channel gains. The solution for optimizing power allocation for DF relaying is also provided, as well as an asymptotically optimal solution for AF relaying. Numerical results are provided to demonstrate substantial performance gain of the jointly optimal solution over some suboptimal alternatives. It is also observed that more gain is obtained from optimal channel pairing than optimal power allocation through judiciously exploiting the variation among multiple channels. Impact of the variation of channel gain, the number of channels, and the number of hops on the performance gain is also studied through numerical examples.

## I. INTRODUCTION

T HE emerging next-generation wireless systems adopt a multichannel relaying architecture for broadband access and coverage improvement [1], [2]. As opposed to a narrowband single-channel relay, a multichannel relay has access to multiple channels, e.g., different frequency channels or subcarriers in an orthogonal frequency division multiplexing (OFDM)

Manuscript received December 19, 2010; revised May 16, 2011; accepted June 22, 2011. Date of publication July 12, 2011; date of current version September 14, 2011. The associate editor coordinating the review of this manuscript and approving it for publication was Dr. Martin Schubert. This work was supported by Natural Sciences and Engineering Research Council (NSERC) of Canada under Discovery Grants and under the Strategic Project Grant programs. The material in this paper was presented at the International ICST Conference on Heterogeneous Networking for Quality, Reliability, Security and Robustness (QShine), Houston, TX, November 2010.

M. Hajiaghayi and B. Liang are with Department of Electrical and Computer Engineering, University of Toronto, Canada (e-mail: mahdih@comm.utoronto.ca; liang@comm.utoronto.ca).

M. Dong is with the Faculty of Engineering and Applied Science, University of Ontario Institute of Technology, Canada (e-mail: min.dong@uoit.ca).

Color versions of one or more of the figures in this paper are available online at http://ieeexplore.ieee.org.

Digital Object Identifier 10.1109/TSP.2011.2161475

system. It may receive a signal from one channel and transmit a processed version of the signal on a different channel. This *multichannel relaying capability* can be exploited to choose forwarding channel adaptively for the incoming signals, taking advantage of the diverse strength of different channels.

In this work, we address the general problem of channel selection and power allocation strategies at multichannel capable relays to forward data in a multihop relaying network. This problem involves two issues: 1) channel pairing (CP): the pairing of incoming and outgoing channels at each relay; 2) power allocation (PA): the determination of power used to transmit signals on these channels. In general, for multihop relaying, there is strong correlation between CP and PA. Intuitively, to maximize the source-destination performance, the choice of CP at each relay would affect the choices of CP at other relays, which further depends on the specific PA scheme used. The optimal system performance requires joint consideration of CP and PA. Our goal is to maximize the end-to-end data rate in a multihop relaying network.

One may view a CP scheme at each relay as a routing scheme embedded in the network router. However, despite bearing some resemblance, the CP problem differs from the conventional multichannel routing problem: For channel pairing, the total cost of two paired incoming and outgoing links is not additive as it is typically assumed in the routing case. Furthermore, the cost of each link cannot be independently defined in CP. The source-destination achievable data rate is dictated by the end-to-end signal-to-noise ratio (SNR), which is a nonlinear function of the channel gain and power used on each link.

### A. Contributions

In this paper, we present a comprehensive solution for jointly optimizing CP and PA to maximize the source-destination data rate in a multichannel multihop relay network. The main results in our work are summarized as follows:

- Given fixed power allocation, the sorted-SNR CP scheme is shown to be optimal in multihop relaying. Specifically, CP can be separated into individual pairing problems at each relay, where the relay matches the incoming channels to the outgoing channels in the order of SNRs seen over these channels.
- The problem of joint CP and PA optimization can be decomposed into two separate problems which can be solved independently: first CP optimization, and then PA optimization. The decoupling of CP and PA optimization significantly reduces the problem search space and reduces the complexity of optimal solution. This separation principle holds for both amplify-and-forward (AF) and decode-and-forward (DF) relaying strategies, and for either total or individual power constraints imposed on the transmitting nodes.

- In joint CP and PA optimization, the optimal CP is shown to be decoupled into per relay CP. The channels at two consecutive hops are optimally paired according to their channel gain order, without the need for knowledge of power allocation on each channel. This allows simple distributed relay implementation for optimal operation, as well as easily adapting to the network topology changes.

- The solution for PA optimization in a multihop setting is proposed for both AF and DF relaying. For DF relaying, we develop an algorithm through a dual-decomposition approach, where we are able to obtain the semiclosed-form PA expression. In addition to depicting power distribution across channels at an individual node, the PA expression allows us to characterize the interaction among the nodes for power determination on a multihop path. For AF relaying, an asymptotically optimal PA solution is provided.

The above results are obtained first for the single-destination scenario. We further extend these results to the multidestination scenario for sum-rate maximization, where we show that the last relay should assign each outgoing channel to a user with the strongest channel gain, and the other relays act the same as in the single-destination scenario.

The separation of joint CP and PA has been established for dual-hop (i.e., single-relay) DF relaying in prior work under total power constraints [3][1] and individual power constraints [4]. It is somewhat surprising that such separation property is preserved in the general multihop relaying. In fact, the generalization from the dual-hop case to the multihop case is nontrivial. For the latter, in addition to being a function of power allocation, the pairing at each relay along the hops needs to be optimized jointly, adding an additional dimension for the optimization problem. Intuitively, to maximize the source-destination rate, the choice of CP at each relay would affect the choices of CP at other relays, which also depend on the specific power allocation scheme used. Therefore, it is not apparent that the optimal CP can be decomposed into independent pairing problems at each relay, or that CP and PA can be separately considered. Besides, the two different techniques used in [3], [4] to show the separation result are complicated. They cannot be simply extended and applied to the multihop case. Instead, we develop a new approach to attack the problem that leads to the separation principle of joint CP and PA for *both AF and DF* relaying in a general multihop setting, under either total or individual power constraints. Our approach provides a rigorous and direct way in proving the separation result.

We further provide numerical studies on the performance of jointly optimal CP and PA scheme and compare it with those of other alternatives for multichannel multihop relaying. We will show that, although both CP and PA improves the performance, the optimal CP is more crucial than the optimal PA. In other words, a major portion of the gain comes from the optimal CP. In addition, we will see that uniform PA with optimal CP achieves near-optimal performance even at moderately high SNR, for AF relaying. This significantly simplifies the PA implementation, without the need of centralized channel information for either CP or PA. The gain by the optimal CP widens with a higher level of channel gain variation across channels, or a larger number of channels, indicating that these factors can be judiciously exploited through CP. The optimal PA, on the other hand, is insensitive to these changes. Finally, we will also show that the gain of jointly optimal CP and PA becomes more pronounced with an increasing number of hops.

## B. Related Work

For an OFDM system as a typical example of multichannel systems, the concept of CP was first introduced independently in [5] and [6] for a dual-hop AF relaying system where heuristic algorithms for pairing based on the order of channel quality were proposed. For relaying without the direct source-destination link available, [5] used integer programming to find the optimal pairing that maximizes the sum SNR. From a system-design perspective, the sorted-SNR CP scheme was proposed in [6] and was shown optimal for the noise-free relaying case, under the assumption of uniform power allocation.

These works sparked interests for more research in this area. In the absence of the direct source-destination link, for the practical case of noisy-relay, by using the property of L-superadditivity of the rate function, the authors of [7] proved that the sorted-SNR CP still remains optimal for sum-rate maximization in dual-hop AF relaying OFDM system. Subsequently, it was further proved in [8], through a different approach, that the sorted-SNR CP scheme is optimal for both AF and DF relaying in the same setup. When the direct source-destination link is available, [9] presented two suboptimal CP schemes. For the same setup, a low complexity optimal CP scheme was later established in [10] for dual-hop AF relaying, and the effect of direct path on the optimal pairing was characterized. In addition, it was shown in [10] that, under certain conditions on relay power amplification, among all possible linear processing at the relay, the channel pairing is optimal.

The related problem of optimal PA for a dual-hop OFDM system was studied by many [11]–[13] for different relay strategies and power constraints. The problem of jointly optimizing CP and PA was studied in a dual-hop OFDM system for AF and DF relaying in [14] and [15], respectively, where the direct source-destination link was assumed available. The joint optimization problems were formulated as mixed integer programs and solved in the Lagrangian dual domain. Exact optimality under arbitrary number of channels was not established. Instead, by adopting the time-sharing argument [16] in their systems, the proposed solutions were shown to be optimal in the limiting case as the number of channels approaches infinity.

Without the direct source-destination link, jointly optimizing CP and PA for DF relaying in a dual-hop OFDM system was investigated in [3] and [4], where [3] assumed a total power constraint shared between the source and the relay, and [4] considered individual power constraints separately imposed on the source and the relay. In both cases, two-step separate CP and PA schemes were proposed and then proved to achieve the jointly optimal solution. For this dual-hop setup, it was shown that the optimal CP scheme is the one that maps the channels solely based on their channel gains independent of the optimal PA solution.

Similar studies on the problem of CP and PA in *dual-hop AF* relaying or *multihop* relaying have been scarce. The authors of [17] proposed an adaptive PA algorithm to maximize the end-to-end rate under the total power constraint in a multihop OFDM relaying system. For a similar network with DF relaying, [18] studied the problem of joint power and time allocation under the long-term total power constraint to maximize the end-to-end rate. Furthermore, in [17], the idea of using CP to further enhance the performance was mentioned in addition to PA. However, no claim was provided on the optimality of the pairing scheme under the influence of PA. The optimal joint CP and PA solution remained unknown.

---

[1] A flaw in the proof was later found in the paper. However, an email correspondence with the authors confirmed that, it can be corrected to show the same result.

### C. Organization

The rest of this paper is organized as follows. In Section II, we present the system model and joint optimization formulation. In Section III, given a fixed PA solution, we provide the optimal CP scheme based on the sorted SNR for both AF and DF strategies. The joint optimization problem of CP and PA is considered in Section IV, where the separation principle between CP and PA optimization is established. The optimal PA solution is then discussed in Section V for multihop relaying under both total and individual power constraints. In Section VI, we further extend the joint optimization and separation results to the multidestination scenario. The numerical study are provided in Section VII, and finally we conclude the paper in Section VIII.

## II. SYSTEM MODEL AND PROBLEM STATEMENT

We mainly focus on an $M$-hop relay network where a source node communicates with a destination node via $(M - 1)$ intermediate relay nodes as illustrated in Fig. 1. Extension to the multidestination scenario turns out to be direct and is presented in Section VI. For broadband communication between the nodes, the frequency bandwidth is split into multiple subbands for data transmission. A practical system with such an approach is the OFDM system where the bandwidth is divided into $N$ equal-bandwidth channels. We denote by $h_{m,n}$, for $m = 1, \ldots, M$ and $n = 1, \ldots, N$, the channel response on channel $n$ over hop $m$. The additive noise at hop $m$ is modeled as an i.i.d. zero mean Gaussian random variable with variance $\sigma_m^2$. We define $a_{m,n} \triangleq \frac{|h_{m,n}|^2}{\sigma_m^2}$ as the *normalized* channel gain against the noise power over channel $n$ of hop $m$. In the rest of presentation, we simply refer to it as channel gain without causing confusion. We make the common assumption that the full knowledge of global channel gains is available at a central controller, which determines the optimal CP and PA[2]. We further assume that the destination is out of the transmission zone of the source, and therefore, there is no direct transmission link. For $M$-hop relaying, a transmission from source to destination occupies $M$ equal time slots, one for each hop. In the $m$th slot, $m = 1, \ldots, M$, the $m$th node (the source node if $m = 1$, otherwise the $(m - 1)$th relay node) transmits a data block to the $(m + 1)$th node (the destination node if $m = M$, otherwise the $m$th relay node) on each channel. Our study is constrained to half-duplex transmissions, where the relay nodes cannot send and receive at the same time on the same frequency. However, the transmission of different data blocks in different hops may occur concurrently, depending on the scheduling pattern for spatial reuse of spectrum.

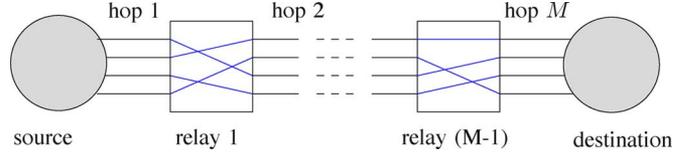

Fig. 1. Illustration of multichannel multihop relaying network with channel pairing.

### A. Relaying Strategies

We consider two types of relaying strategies: AF and DF. In AF relaying, a relay amplifies the data received from an incoming channel and directly forwards it to the next node over an outgoing channel. In DF relaying, a relay attempts to decode the received data from the previous node over each incoming channel and forwards a version of the decoded data on an outgoing channel to the next node. We consider the simple repetition-coding based DF relaying [19], [20], where the relay is required to fully decode the incoming message, re-encodes it with repetition coding, and forwards it to the intended receiver.

### B. Channel Pairing

The relay conducts CP, matching each incoming channel with an outgoing channel. As different channels exhibit various quality, a judicious CP scheme can potentially lead to significant improvement in system spectral efficiency.

We denote path $\mathcal{P}_i = (c(1, i), \ldots, c(M, i))$, where $c(m, i)$ specifies the index of the channel at hop $m$ that belongs to path $\mathcal{P}_i$. For example, $\mathcal{P}_i = (3, 4, 2)$ indicates that path $\mathcal{P}_i$ consists of the third channel at hop 1, the fourth channel at hop 2, and the second channel at hop 3. Once channel pairing is determined at all the relays, the total $N$ disjoint paths $\mathcal{P}_1, \ldots, \mathcal{P}_N$ can be identified from the source to the destination.

### C. Power Allocation

Denote the power allocated to channel $n$ over hop $m$ by $P_{m,n}$. The SNR obtained on this channel is represented by $\gamma_{m,n} = P_{m,n} a_{m,n}$. For each path $\mathcal{P}_i$, let $\tilde{\gamma}_{m,i} \triangleq \gamma_{m,c(m,i)}$ represent the SNR seen over hop $m$ on this path.

Let $\mathbf{P}_i = (P_{1,c(1,i)}, \ldots, P_{M,c(M,i)})$ be the PA vector for all channels along path $\mathcal{P}_i$. The source-destination equivalent SNR of path $\mathcal{P}_i$ is denoted by $\gamma_{SD}(\mathcal{P}_i, \mathbf{P}_i)$. For AF relaying, it is given by [21]

$$\gamma_{SD}^{AF}(\mathcal{P}_i, \mathbf{P}_i) = \left( \prod_{m=1}^{M} \left( 1 + \frac{1}{\tilde{\gamma}_{m,i}} \right) - 1 \right)^{-1} \quad (1)$$

and, in Section V, we will also use its upper bound [21]

$$\gamma_{SD}^{AF}(\mathcal{P}_i, \mathbf{P}_i) \approx \left( \sum_{m=1}^{M} \frac{1}{\tilde{\gamma}_{m,i}} \right)^{-1} \quad (2)$$

whose approximation gap vanishes as the SNR becomes large. For DF relaying, we have

$$\gamma_{SD}^{DF}(\mathcal{P}_i, \mathbf{P}_i) = \min_{m=1,\ldots,M} \tilde{\gamma}_{m,i}. \quad (3)$$

We consider two types of power constraints.

*a) Total Power Constraint:* The power assignment $P_{m,n}$, for $m = 1 \cdots M$ and $n = 1 \cdots N$, must satisfy the following aggregated power constraint:

---

[2]However, we show later that, for joint CP and PA optimization, the CP solution requires only local channel information at each relay, and given the proposed CP solution, a uniform PA scheme without channel information is near optimal even at moderately high SNR for AF relaying.

$$\sum_{m=1}^{M}\sum_{n=1}^{N}P_{m,n}=P_t. \qquad (4)$$

*b) Individual Power Constraint:* The power assignment $P_{m,n}$, for $n=1,\dots,N$, needs to satisfy the power constraint of the individual node $m$, i.e.

$$\sum_{n=1}^{N}P_{m,n}=P_{mt},\quad m=1,\dots,M \qquad (5)$$

where $P_{mt}$ denotes the maximum allowable power at node $m$.

### D. Objective

Our goal is to design a jointly optimal CP and PA strategy to maximize the source-destination rate under multihop relaying. The source-destination rate achieved through path $\mathcal{P}_i$ is given by

$$R_{\text{SD}}(\mathcal{P}_i,\mathbf{P}_i)=\frac{1}{F_s}\log_2(1+\gamma_{\text{SD}}(\mathcal{P}_i,\mathbf{P}_i))$$

where $F_s$ is the spatial reuse factor. In multihop relaying that allows concurrent transmissions, $F_s$ takes value between 2 and $M$ (since $F_s\geq 2$ under the half-duplex assumption). The sum rate of all paths determines the total source-destination rate of the system, denoted as $R_t$, i.e.

$$R_t=\sum_{i=1}^{N}R_{\text{SD}}(\mathcal{P}_i,\mathbf{P}_i). \qquad (6)$$

It is a function of both $\{\mathcal{P}_i\}$ and $\{\mathbf{P}_i\}$, which should be jointly optimized

$$\max_{\{\mathcal{P}_i\},\{\mathbf{P}_i\}}\quad R_t \qquad (7)$$
$$\text{subject to}\quad (4)\quad\text{or}\quad(5)$$
$$\mathbf{P}_i\succeq 0,\quad i=1,\dots,N \qquad (8)$$

where $\succeq$ signifies element-wise inequality.

## III. OPTIMAL MULTIHOP CHANNEL PAIRING UNDER FIXED POWER ALLOCATION

In this section, we first consider the case when PA is fixed and given. In this case, the optimization problem in (7) can be rewritten as

$$\max_{\{\mathcal{P}_i\}}\sum_{i=1}^{N}R_{\text{SD}}(\mathcal{P}_i,\mathbf{P}_i) \qquad (9)$$

and the optimal CP $\{\mathcal{P}_i^*\}$ is a function of $\{\mathbf{P}_i\}$. To simplify the notation, in this section we rewrite $R_{\text{SD}}(\mathcal{P}_i)$ and $\gamma_{SD}(\mathcal{P}_i)$ and drop their dependency on $\mathbf{P}_i$ with the understanding that $\{\mathbf{P}_i\}$ is fixed. In the following, we solve (9) to obtain the optimal CP scheme under this fixed PA. We emphasize that here the generalization from the dual-hop case to the multihop case is nontrivial. Intuitively, there is no obvious way to decouple the sequence of pairings at all $(M-1)$ relays. Indeed, the *equivalent* incoming channel from a source to a relay and the *equivalent* outgoing channel from that relay to the destination depend on how the channels are paired over multiple hops. However, we will show that the optimal CP solution over multiple hops can in fact be decomposed into $(M-1)$ independent CP problems, where the mapping of incoming and outgoing channels at each relay is only based on the sorted SNR over those channels, and therefore can be performed individually per hop.

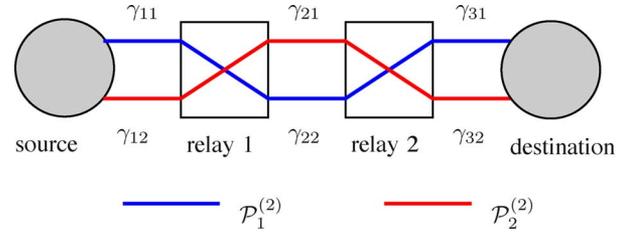

Fig. 2. Three-hop relay with two channels.

In the following, we first establish the optimality of the sorted-SNR CP scheme for the case of $M=3$ and $N=2$, and then we extend the result to arbitrary $M$ and $N$.

### A. Optimal Channel Pairing for Three-Hop Relaying

*1) Two-Channel Case ($N=2$):* We first consider a three-hop relaying network with two channels, as depicted in Fig. 2. Without loss of generality, we assume channel 1 exhibits equal or larger SNR than channel 2 over all the three hops, i.e.

$$\gamma_{m,1}\geq\gamma_{m,2},\text{ for } m=1,2,3. \qquad (10)$$

The optimal CP scheme for this case is given in Proposition 1.

*Proposition 1:* For $M=3$ and $N=2$, the solution to (9) is the sorted-SNR CP scheme performed on each relay, i.e., $\{\mathcal{P}_i^*\}=\{(1,1,1),(2,2,2)\}$ under (10).

*Proof:* The proof essentially exams possible path selections and shows that the sorted-SNR per relay provides the highest source-destination sum rate for both AF and DF relaying. See Appendix A for details. ∎

*2) Multichannel Case ($N\geq 2$):* Here, we provide an argument to extend the result in Proposition 1 to a system with an arbitrary number of channels.

*Proposition 2:* For $M=3$ and $N\geq 2$, the solution to (9) is the sorted-SNR CP scheme performed on each relay.

*Proof:* Suppose the optimal pairing does not follow the pairing rule of sorted SNR. There is at least one relay (say, Relay 2) that has two pairs of incoming and outgoing channels that are mismatched according to their SNR. That is, there exist two channels $i_1$ and $i_2$ over hop 2, and two channels $j_1$ and $j_2$ over hop 3 that are respectively paired with each other while $\gamma_{2,i_1}<\gamma_{2,i_2}$ and $\gamma_{3,j_1}>\gamma_{3,j_2}$. Note that these two channel pairs belong to two disjoint source-destination paths that can be regarded as a 2-channel relay system. From Proposition 1, we know that pairing channels $i_1$ with $j_2$ and $i_2$ with $j_1$ at relay 2 achieves a higher rate than the existing pairing over these two paths. Hence, by switching to this new pairing while keeping the other paths the same, we could increase the total rate. This contradicts our assumption on the optimality of a nonsorted SNR CP scheme.

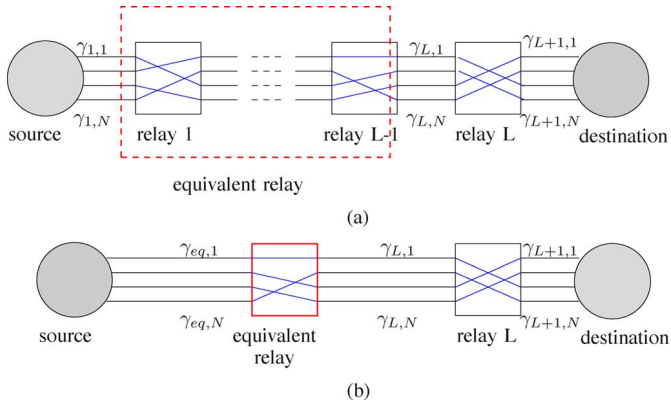

Fig. 3.   Converting an $(L+1)$-hop relaying to an equivalent 3-hop relaying. (a) An $(L+1)$-hop relaying network. (b) An equivalent 3-hop relaying network.

Hence, there is no better scheme than sorted-SNR CP to obtain the maximum sum rate.   ∎

### B. Optimal Channel Pairing for Multihop Relaying

Building on Proposition 2, we next extend the result for 3-hop relaying to a relaying network with an arbitrary number of hops $(M \geq 3)$ in the following proposition.

*Proposition 3:* The solution to (9) is the sorted-SNR CP scheme individually performed at each relay.

*Proof:* We prove by induction. It is shown in Proposition 2 that the sorted-SNR CP is optimal for $M = 3$. Suppose the claim holds for $M \leq L$. Now consider $M = L + 1$ as shown in Fig. 3(a). Let $\gamma_{eq,n}$ be the received SNR from the source to relay $L - 1$ over the $n$th incoming channel of that relay. We establish $N$ equivalent channels between the source and relay $L - 1$, with SNR over the $n$th channel as $\gamma_{eq,n}$. Then, the $(L+1)$-hop relaying network can be converted to a 3-hop network, with an equivalent relay whose incoming channels have SNR $\{\gamma_{eq,n}\}$ and outgoing channels remain the same as those of relay $L - 1$, as shown in Fig. 3(b). Hence, from Proposition 2, the optimal CP is the one where $\{\gamma_{eq,n}\}$ and $\{\gamma_{L,n}\}$ are sorted and paired at this equivalent relay, and $\{\gamma_{L,n}\}$ and $\{\gamma_{L+1,n}\}$ are sorted and paired at relay $L$. Note that the sorted-SNR pairing at relay $L$ is independent of how the channels are paired at the other relays.

Next, ignore relay $L$ and replace it by equivalent channels from relay $L - 1$ to the destination. We now have a $L$-hop network. From the induction hypothesis, the sorted-SNR CP is optimal. In particular, the incoming and outgoing channels at each of relays $1, 2, \ldots, L - 2$ are sorted by their SNR and paired. Since the SNRs $\{\gamma_{eq,n}\}$ at the equivalent relay are computed by applying (1) to (3) over these *sorted and paired* channels from the source to relay $L - 1$, it is not difficult to see that $\{\gamma_{L-1,n}\}$ and $\{\gamma_{eq,n}\}$ are ordered in the same way. Therefore, sorting and pairing $\{\gamma_{eq,n}\}$ and $\{\gamma_{L,n}\}$ at the equivalent relay is the same as sorting and pairing $\{\gamma_{L-1,n}\}$ and $\{\gamma_{L,n}\}$ at relay $L - 1$. Thus, we conclude that at each of relay $1, \ldots, L$, the incoming and outgoing channels are sorted and paired in order of their SNR. ∎

The significance of Proposition 3 is that the optimal CP for $M$-hop relaying is decoupled into $(M - 1)$ individual pairing schemes at each relay, each solely based on the SNR of incoming and outgoing channels. This decoupling not only reduces the pairing complexity, but also reveals the distributed nature of optimal CP among multiple relays, thus allowing simple implementation that can easily adapt to network topology changes.

*Remark:* We point out that the existing result of optimal CP strategy for dual-hop relaying is not sufficient for the induction to prove Proposition 3. Notice that, in the proof, an $M$-hop network $(M > 3)$ was transformed into an equivalent 3-hop network. Reducing a 3-hop network to a dual-hop network would require combining relay nodes with either the source or the destination to form an equivalent node and equivalent channel gain. The dual-hop result can only be applied to pairing with the *equivalent channels*, but is not sufficient to show the actual physical channel should follow the same pairing strategy. Therefore, Proposition 2 is necessary as the basis to prove the general $M$-hop case.

In addition, in [7], the L-superadditivity property [22] is used to show that the sorted-SNR CP is optimal in dual-hop AF relaying for sum-rate maximization. That is, if the source-destination rate over each path can be shown to be L-superadditive, it follows that sorted-SNR pairing is optimal. However, L-superadditivity does not hold for the rate function in general multihop relaying, where the source-destination rate is a higher dimensional function defined on $\mathbb{R}^M$ with respect to $\gamma_{1,n}, \ldots, \gamma_{M,n}$, for a given $n$. Thus, a similar proof for the optimality of sorted-SNR in the dual-hop case is not available to the general multihop case.

## IV. JOINTLY OPTIMAL CHANNEL PAIRING AND POWER ALLOCATION: A SEPARATION PRINCIPLE

So far, given a fixed PA scheme, we have found that the optimal CP scheme for (9) is SNR based, which depends on the transmission power allocated to each channel. We next present the solution for (7) by jointly optimizing CP and PA.

The apparent coupling of CP and PA makes a direct exhaustive search for the jointly optimal solution prohibitively complex. Instead, we will show that the joint optimization problem can be decoupled into two separate CP and PA subproblems. Specifically, we prove that the jointly optimal solution is obtained by pairing channels based on the order of their *channel gains* (normalized against the noise power), followed by optimal PA over the paired channels. This separation principle holds for a variety of scenarios, including AF and DF relaying under either total or individual power constraints.

Our argument for the separation principle is briefly summarized as follows. We first show that, at a global optimum, the channel with a higher channel gain exhibits a larger SNR. This relation reveals that the SNR-based ordering of channels is the same as the one based on channel gain. Hence, we conclude that the sorted CP scheme based on channel gain is optimal when PA is also optimized.

### A. Ordering Equivalence at Optimality

Let $\gamma_{m,n}^*$ be the received SNR under the optimal PA solution $\{P_n^*\}$ for hop $m$ and channel $n$. For both total and individual power constraints, the following proposition establishes the equivalence between channel-gain ordering and SNR-based ordering at the optimality.

*Proposition 4:* In the optimal CP and PA solution for (7), at each hop, the channel with better channel gain also provides a higher received SNR, i.e., $a_{m,i} \geq a_{m,j}$ implies $\gamma_{m,i}^* \geq \gamma_{m,j}^*$, for $m = 1, \ldots, M; i, j \in \{1, \ldots, N\}$, and $i \neq j$.

*Proof:* We first provide a proof for a multihop system consisting of two channels. We then explain how it can be extended to a system with an arbitrary number of channels.

*a) $N = 2$:* We prove the proposition by contradiction. Let $\mathcal{P}_1$ and $\mathcal{P}_2$ represent the two disjoint source-destination paths corresponding to the optimal CP scheme. Consider any hop $m$ along these paths. Without loss of generality, let channel 1 belong to $\mathcal{P}_2$, channel 2 belong to $\mathcal{P}_1$, and $a_{m,1} \geq a_{m,2}$. Suppose at optimality $\gamma_{m,1}^* < \gamma_{m,2}^*$, i.e., $P_{m,2}^* a_{m,2} > P_{m,1}^* a_{m,1}$, where $P_{m,1}^*$ and $P_{m,2}^*$ are the power allocated to channels 1 and 2, respectively. Let $P_{mt} = P_{m,1}^* + P_{m,2}^*$.

Consider the following alternate allocation of power between channels 1 and 2 over hop $m$

$$\hat{P}_{m,1} = \frac{a_{m,2}}{a_{m,1}} P_{m,2}^*$$
$$\hat{P}_{m,2} = \frac{a_{m,1}}{a_{m,2}} P_{m,1}^*. \tag{11}$$

We further swap the two channels so that channel 1 belongs to path $\mathcal{P}_1$ and channel 2 belongs to path $\mathcal{P}_2$. Since $\hat{P}_{m,1} a_{m,1} = P_{m,2}^* a_{m,2}$ and $\hat{P}_{m,2} a_{m,2} = P_{m,1}^* a_{m,1}$, the above procedure of power re-allocation and channel swapping does not change the end-to-end rate.

$$
\begin{aligned}
\hat{P}_{m,1} + \hat{P}_{m,2} &= \frac{a_{m,2}}{a_{m,1}}(P_{mt} - P_{m,1}^*) + \frac{a_{m,1}}{a_{m,2}} P_{m,1}^* \\
&= \frac{a_{m,2}}{a_{m,1}} P_{mt} + \frac{(a_{m,1})^2 - (a_{m,2})^2}{a_{m,1} a_{m,2}} P_{m,1}^* \\
&< \frac{a_{m,2}}{a_{m,1}} P_{mt} \\
&\quad + \frac{(a_{m,1})^2 - (a_{m,2})^2}{a_{m,1} a_{m,2}} \frac{a_{m,2}}{a_{m,1} + a_{m,2}} P_{mt} \\
&= P_{mt}
\end{aligned} \tag{12}
$$

where inequality (12) is obtained from our assumption that $P_{m,1}^* a_{m,1} < P_{m,2}^* a_{m,2}$, which can be rewritten as $P_{m,1}^* < \frac{a_{m,2}}{a_{m,1} + a_{m,2}} P_{mt}$, and that $a_{m,1} \geq a_{m,2}$. This contradicts our initial assumption that the original PA is globally optimal.

*b) $N > 2$:* A similar proof by contradiction as it was used in Section III–A–2 for Proposition 2 can be applied to generalize the above result to $N > 2$. For an $N$-channel relay system with $N > 2$, suppose the optimal CP scheme follows the pairing rule of the sorted CP based only on SNR gain and not channel gain. As a result, there is at least one hop over which, between two channels, the channel with better channel gain demonstrates a lower SNR. These two channels essentially belong to the two source-destination paths that can be considered as a 2-channel relay system. From the above, we know that by just swapping these two channels and applying the alternate allocation of power in (11), the sum power is reduced while maintaining the same rate. This leads to a contradiction of our early assumption on the optimality of the sorted CP not being conducted based on the channel gain. ∎

### B. Separation Principle

*Proposition 5:* The joint optimization of CP and PA in (7) can be separated into the following two steps:

1) Obtain the optimal CP $\{\mathcal{P}_i^*\}$. The optimal CP $\{\mathcal{P}_i^*\}$ is independent of $\{\mathbf{P}_i^*\}$ and is performed individually at each relay in the order of sorted channel gain.

2) Obtain the optimal PA $\{\mathbf{P}_i^*\}$ under the optimal CP $\{\mathcal{P}_i^*\}$

$$\{\mathbf{P}_i^*\} = \arg\max_{\{\mathbf{P}_i\}} \sum_{i=1}^N R_{\mathrm{SD}}(\mathcal{P}_i^*, \mathbf{P}_i)$$
$$\text{subject to (4) or (5).} \tag{13}$$

*Proof:* From Proposition 3, with optimal PA $\{\mathbf{P}_i^*\}$, the sorted-SNR CP gives the optimal $\{\mathcal{P}_i^*\}$. From Proposition 4, at optimality, the sorted-SNR CP is equivalent to sorting channel gains, which does not require the knowledge of $\{\mathbf{P}_i^*\}$. The optimal $\{\mathbf{P}_i^*\}$ then can be obtained under the optimal CP, and we have the separation principle. ∎

Decoupling the CP strategy from PA strategy significantly reduces the problem search space. In addition, the optimal CP strategy in the presence of multiple hops is further decoupled into independent sorting problems at each hop, which only depends on the channel gain on the incoming and outgoing channels. The complexity of the optimal CP strategy for each hop is that of sorting channel gain, which is $O(N \log N)$. Therefore, the total complexity of the joint CP and PA optimization is $O(MN \log N)$ in addition to the complexity of PA optimization.

## V. Optimal Power Allocation for Multihop Relaying

So far we have obtained the optimal CP at all relays. We next find the optimal PA solution for a given CP scheme as in (13). With the channels paired at each relay, the system can be viewed as a regular multihop system. Without loss of generality, we assume the channel gains at each hop are in descending order according to their channel index, i.e., $a_{m,1} \geq a_{m,2} \geq \cdots \geq a_{m,N}$, for $m = 1, \ldots, M$. From Proposition 5, the channels with the same index are paired, and a path with the optimal CP consists of all the same channel index, i.e., $\mathcal{P}_i^* = (i, \ldots, i)$. In the following, we consider the PA optimization problem for total power and individual power constraints separately.

### A. Total Power Constraint

The optimal PA solution with a total power constraint for a multihop relaying OFDM system was obtained in [17]. The results can be directly applied here. We briefly state the solution for completeness.

The PA optimization problem in (13) with a total power constraint has the classical water-filling solution

$$P_i^* = \left[ \frac{1}{\lambda} - \frac{1}{a_{eq,i}} \right]^+ \quad \text{for } i = 1, \ldots, N \tag{14}$$

where $[x]^+ = \max(x, 0)$. The Lagrange multiplier $\lambda$ is chosen such that the power constraint in (4) is met, and $a_{eq,i}$ is an equivalent channel gain over the path $\mathcal{P}_i^*$.

For DF relaying, the equivalent channel gain, denoted as $a_{eq,i}^{\mathrm{DF}}$, is given by [17]

$$a_{eq,i}^{\mathrm{DF}} = \left( \sum_{m=1}^M \frac{1}{a_{m,i}} \right)^{-1}, \quad i = 1, \ldots, N. \tag{15}$$

In other words, the equivalent channel gain is $N$ times the harmonic mean of the channel gain over each hop. It is obtained following the fact that, to maximize the source-destination rate on one path, the total power allocated to the path must be shared among the channels on this path such that all channels exhibit the same SNR. The power allocated to each transmitting node on path $\mathcal{P}_i^*$ is given by

$$P_{m,i}^* = \frac{P_i^*}{a_{m,i}\sum_{m'=1}^M \frac{1}{a_{m',i}}}. \tag{16}$$

For AF relaying, the exact expression for equivalent channel gain on path $\mathcal{P}_i^*$ is difficult to obtain. However, its upper bound approximation can be expressed as [17]

$$a_{\mathrm{eq},i}^{\mathrm{AF}} \approx \left(\sum_{m=1}^M \frac{1}{\sqrt{a_{m,i}}}\right)^{-2}, \quad i = 1,\ldots,N. \tag{17}$$

In this case, the equivalent channel amplitude (normalized against noise standard deviation) is $N$ times the harmonic mean of the channel amplitude over each hop. It is obtained using the upper bound approximation of equivalent SNR in (2) over a path. The power allocated to each transmitting node on path $\mathcal{P}_i^*$ is given by

$$P_{m,i}^* = \frac{P_i^*}{\sqrt{a_{m,i}}\sum_{m'=1}^M \frac{1}{\sqrt{a_{m',i}}}}. \tag{18}$$

The PA solution in (14) requires global channel gain information and therefore needs to be implemented in a centralized fashion.

### B. Individual Power Constraint

For DF relaying, the source-destination sum rate in (6) reduces to

$$R_t^{\mathrm{DF}} = \frac{1}{F_s}\sum_{n=1}^N \min_{m=1,\ldots,M} \log_2(1 + P_{m,n}a_{m,n}). \tag{19}$$

Maximizing (19) over $\{P_{m,n}\}$ under individual power constraints in (5) can be cast into the following optimization problem using a set of auxiliary variables $\mathbf{r} = [r_1,\ldots,r_N]^T$:

$$\max_{\mathbf{r},\mathbf{P}} \frac{1}{F_s}\sum_{n=1}^N r_n \tag{20}$$

$$\text{subject to } i) \quad r_n \leq \log_2(1 + P_{m,n}a_{m,n})$$
$$m = 1,\ldots,M, \quad n = 1,\ldots,N;$$

$$ii) \quad \sum_{n=1}^N P_{m,n} \leq P_{mt}, \quad m = 1,\ldots,M;$$

$$iii) \quad \mathbf{P} \succeq 0$$

where $\mathbf{P} \triangleq [P_{m,n}]_{M\times N}$. Since the objective function is linear, and all the constraints are convex, the optimization problem in (20) is convex. For such a problem, Slater's condition holds [23], and the duality gap is zero. Thus, (20) can be solved in the Lagrangian dual domain. Since the spatial reuse factor $F_s$ is a constant, we drop it for simplicity without affecting the optimization problem. Consider the Lagrange function for (20),

$$\mathcal{L}(\mathbf{P},\mathbf{r},\boldsymbol{\mu},\boldsymbol{\lambda}) = \sum_{n=1}^N r_n$$
$$- \sum_{n=1}^N \sum_{m=1}^M \mu_{m,n}(r_n - \log_2(1 + P_{m,n}a_{m,n}))$$

$$- \sum_{m=1}^M \lambda_m \left(\sum_{n=1}^N P_{m,n} - P_{mt}\right) \tag{21}$$

where $\boldsymbol{\mu} \triangleq [\mu_{m,n}]_{M\times N}$ with $\mu_{m,n}$ being the Lagrange multiplier corresponding to constraint $(i)$ in (20), and $\boldsymbol{\lambda} = [\lambda_1,\ldots,\lambda_M]^T$ with $\lambda_m$ being the Lagrange multiplier associated with power constraint in $(iii)$ in (20). The dual function is given by

$$g(\boldsymbol{\lambda},\boldsymbol{\mu}) = \max_{\mathbf{r},\mathbf{P}} \mathcal{L}(\mathbf{P},\mathbf{r},\boldsymbol{\mu},\boldsymbol{\lambda})$$
$$\text{subject to } \mathbf{P} \succeq 0. \tag{22}$$

Optimizing (21) over $\mathbf{r}$ for given $\mathbf{P}$, $\boldsymbol{\mu}$ and $\boldsymbol{\lambda}$ yields

$$\sum_{m=1}^M \mu_{m,n} = 1, \quad \text{for} \quad n = 1,\ldots,N. \tag{23}$$

Substituting this into $\mathcal{L}(\mathbf{P},\mathbf{r},\boldsymbol{\mu},\boldsymbol{\lambda})$, we obtain

$$\mathcal{L}(\mathbf{P},\mathbf{r},\boldsymbol{\mu},\boldsymbol{\lambda})$$
$$= \sum_{n=1}^N \sum_{m=1}^M (\mu_{m,n}\log_2(1 + P_{m,n}a_{m,n}) - \lambda_m P_{m,n})$$
$$+ \sum_{m=1}^M \lambda_m P_{mt}. \tag{24}$$

It is clear that the dual function $g(\boldsymbol{\mu},\boldsymbol{\lambda})$ obtained by maximizing (24) can be decomposed into $NM$ subproblems

$$g(\boldsymbol{\mu},\boldsymbol{\lambda}) = \sum_{n=1}^N \sum_{m=1}^M g_{mn}(\mu_{mn},\lambda_m) + \sum_{m=1}^M \lambda_m P_{mt},$$

with

$$g_{mn}(\mu_{mn},\lambda_m) = \max_{P_{m,n}} \quad \mathcal{L}_{mn}(P_{m,n},\mu_{m,n},\lambda_m)$$
$$\text{subject to} \quad P_{m,n} \geq 0 \tag{25}$$

where

$$\mathcal{L}_{mn}(P_{m,n},\mu_{m,n},\lambda_m)$$
$$= \mu_{m,n}\log_2(1 + P_{m,n}a_{m,n}) - \lambda_m P_{m,n}$$

for $m = 1,\ldots,M; n = 1,\ldots,N$. By applying KKT conditions [23] to (25), the optimal power allocation $P_{m,n}^*$, as a function of $\mu_{m,n}$ and $\lambda_m$, is derived as

$$P_{m,n}^* = \left[\frac{\mu_{m,n}}{\lambda_m} - \frac{1}{a_{m,n}}\right]^+ \tag{26}$$

for $n = 1,\ldots,N$ and $m = 1,\ldots,M$, where $\lambda_m$ is chosen to meet the power constraint in (20).

Finally, the optimization problem in (20) is equivalent to the dual problem

$$\min_{\boldsymbol{\mu},\boldsymbol{\lambda}} g(\boldsymbol{\mu},\boldsymbol{\lambda})$$
$$\text{subject to} \quad \boldsymbol{\mu} \succeq 0, \boldsymbol{\lambda} \succeq 0;$$
$$\sum_{m=1}^M \mu_{m,n} = 1, \quad \text{for} \quad n = 1,\ldots,N. \tag{27}$$

This dual problem can be efficiently solved by using the *projected* subgradient method [24]. Analogous to a common subgradient method, a sequence of Lagrange multipliers is generated which converges to the optimal $\boldsymbol{\lambda}^*$ and $\boldsymbol{\mu}^*$ minimizing $g(\boldsymbol{\mu}, \boldsymbol{\lambda})$. This convergence is achieved provided that a suitable step size is chosen at each iteration [24]. The difference between projected and normal subgradient methods lies in having an extra constraint $\sum_{m=1}^{M} \mu_{m,n} = 1$. To satisfy this constraint, at each iteration, the projected subgradient method projects the columns of $\boldsymbol{\mu}$ (obtained by subgradient method) onto a unit space to attain a set of feasible multipliers. At each iteration, a subgradient of $g(\boldsymbol{\mu}, \boldsymbol{\lambda})$ at the current values of $\mu_{m,n}$ and $\lambda_m$ is required. Let $[\boldsymbol{\theta}_{\boldsymbol{\mu}}, \boldsymbol{\theta}_{\boldsymbol{\lambda}}]^T$ denote the subgradient, where $\boldsymbol{\theta}_{\boldsymbol{\mu}} = [\theta_{\mu_{1,1}}, \ldots, \theta_{\mu_{M,N}}]^T$ and $\boldsymbol{\theta}_{\boldsymbol{\lambda}} = [\theta_{\lambda_1}, \ldots, \theta_{\lambda_M}]^T$. It is obtained from (24) as

$$\theta_{\mu_{m,n}} = \log_2(1 + P_{m,n}^* a_{m,n}) \qquad (28)$$

for $m = 1, \ldots, M$ and $n = 1, \ldots, N$, and

$$\theta_{\lambda_m} = P_{m,n} - \sum_{n=1}^{N} P_{m,n}^*$$

for $m = 1, \ldots, M$, where $P_{m,n}^*$ is obtained from (26).

For completeness, we summarize the projected subgradient algorithm for solving the dual problem.
1) Initialize $\boldsymbol{\lambda}^{(0)}$ and $\boldsymbol{\mu}^{(0)}$.
2) Given $\lambda_m^{(l)}$ and $\mu_{m,n}^{(l)}$, obtain the optimal values of $P_{m,n}^*$ in (26) for all $m$ and $n$.
3) Update $\boldsymbol{\lambda}^{(l)}$ through

$$\lambda_m^{(l+1)} = \left[ \lambda_m^{(l)} - \theta_{\lambda_m} \nu_\lambda^{(l)} \right]^+$$

for $m = 1, \ldots, M$. Similarly, update $\mu_{m,n}^{(l)}$ followed by unitary space projection, i.e.

$$\mu_{m,n}^{(l+1)} = \frac{\hat{\mu}_{m,n}^{(l+1)}}{\sum_{j=1}^{M} \hat{\mu}_{j,n}^{(l+1)}} \qquad (29)$$

where

$$\hat{\mu}_{m,n}^{(l+1)} = \left[ \mu_{m,n}^{(l)} - \theta_{\mu_{m,n}} \nu_\mu^{(l)} \right]^+ \qquad (30)$$

for $m = 1, \ldots, M$; and $n = 1, \ldots, N$. $\nu_\lambda^{(l)}$ and $\nu_\mu^{(l)}$ are the step sizes at the $l$th iteration for multipliers $\mu$ and $\lambda$, respectively.
4) Let $l = l + 1$; repeat from Step 2 until convergence.

With the optimal $\boldsymbol{\lambda}^*$ and $\boldsymbol{\mu}^*$, the optimal power solution $\mathbf{P}^*$ is determined as in (26), i.e., for $m = 1, \ldots, M$ and $n = 1, \ldots, N$,

$$P_{m,n}^* = \left[ \frac{\mu_{m,n}^*}{\alpha \lambda_m^*} - \frac{1}{a_{m,n}} \right]^+ \qquad (31)$$

where $\mu_{m,n}^*$ satisfies the constraint (23), and at the same time, $\mu_{m,n}^*$ and $\lambda_m^*$ are chosen so that the individual power constraints in constraint (ii) of (20) are met.

The expression of $P_{m,n}^*$ in (31) provides some insight on the structure of the optimal PA for multihop DF relaying: For a given $\boldsymbol{\mu}$, the power allocation across channels at each node is individually determined following a scaled version of the water-filling approach based on the channel gain. The scales are determined jointly among different hops to satisfy the condition of $\boldsymbol{\mu}$ in (23). It essentially requires the received SNR $\gamma_{m,n}$ at each hop of the same path to be equal.

We now consider the PA problem for AF relaying. Unlike DF, the achievable source-destination sum rate for AF is not generally concave in $\{P_{m,n}\}$. Therefore, we have a nonconvex optimization problem formulated as

$$\max_{\mathbf{P}} \frac{1}{F_s} \sum_{n=1}^{N} \log_2 \left( 1 + \left[ \prod_{m=1}^{M} \left( 1 + \frac{1}{P_{m,n} a_{m,n}} \right) - 1 \right]^{-1} \right)$$

subject to

$$i) \quad \sum_{n=1}^{N} P_{m,n} \leq P_{mt}, \quad m = 1, \ldots, M$$

$$ii) \quad \mathbf{P} \succeq 0. \qquad (32)$$

To find the PA solution we resort to an upper bound of the sum rate in (32). Based on (2), an upper-bound approximation for the $M$-hop source-destination sum rate is given by

$$R_t^{\text{up}} = \frac{1}{F_s} \sum_{n=1}^{N} \log_2 \left( 1 + \left( \sum_{m=1}^{M} \frac{1}{P_{m,n} a_{m,n}} \right)^{-1} \right). \qquad (33)$$

This upper bound becomes tight as the channel gain $a_{m,n}$ over each channel increases. Therefore, the PA solution obtained using (33) is asymptotically optimal.

*Lemma 1:* $R_t^{\text{up}}$ in (33) is concave with respect to $\{P_{m,n}\}$.

*Proof:* The proof follows from the concavity of (2) with respect to $\{P_{m,n}\}$, which can be shown by considering its Hessian matrix. The details are given in Appendix B. ∎

Given Lemma 1, the optimization of $\{P_{m,n}\}$ to maximize $R_t^{\text{up}}$ is a convex optimization problem, and we can solve it in the Lagrangian dual domain using KKT conditions [23]. Although a closed-form or semiclosed-form solution for $\{P_{m,n}\}$ is difficult to obtain in this case, we can solve it numerically via standard convex optimization tools.

## VI. Extension to Multiple-Destination Case

The system model we consider so far assumes a single pair of source and destination communicating through $M$-hop relaying. In this section, we show that the results in the previous sections can be extended to a multihop multidestination relaying network for sum-rate maximization.

Specifically, we consider a single source node communicating with $K$ users through $M$-hop relaying via $(M-1)$ common relay nodes[3]. In this multidestination system, the last relay conducts CP as well as channel-user assignment. In channel-user assignment, the relay partitions the $N$ outgoing channels into $K$ subsets, assigning one for each user for data forwarding. To maximize the end-to-end sum-rate of $K$ users, the joint optimization now involves CP, channel-user assignment, and PA. Despite the correlation between CP and channel-user assignment, we show that the results for the single source-destination case can be directly extended to the multidestination case. To see this, we notice that all users share the common $(M-1)$-hop relay channels and the channel-user assignment is performed at the last relay for the $M$th hop. Given a channel-user assignment, this multidestination system can be viewed equivalently as a single-destination system, and the results of optimal CP and PA in the previous sections apply. To optimize the channel-user assignment, it is not difficult to see that, for any given pairing in the first $(M-1)$ hops, at the last relay, assigning each outgoing channel to a user who has the strongest channel gain among all $K$ users maximizes the end-to-end rate. This result is summarized in the following proposition.

---

[3]While this model is appropriate to a downlink scenario, the result and proposed solution is applicable to the uplink scenario by swapping the role of source and destination nodes, where the given power constraint concerns the source nodes only in terms of their total transmission power.

*Proposition 6:* For a $M$-hop $K$-user relaying network described above, the jointly optimal solution of CP, channel-user assignment, and PA that maximizes the end-to-end sum-rate is obtained via two steps:

1) Channel-user assignment: In the last hop, the $(M-1)$th relay assigns its $i$th outgoing channel to a user exhibiting the strongest channel gain among $K$ users over that channel;

2) CP and PA: Under the channel-user assignment from step 1, apply the jointly optimal CP and PA solution of the single source-destination case.

We note that the above result for the multidestination scenario holds for both DF and AF relaying, and for total and individual power constraints.

## VII. NUMERICAL RESULTS

In this section, we provide simulation examples to evaluate and compare the performance of the optimal joint CP and PA scheme with that of suboptimal CP and PA alternatives. We study different factors that affect the performance gap under these schemes.

Besides the jointly optimized CP and PA scheme, the following suboptimal schemes are used for comparison: 1) *Uniform PA with CP*: the optimal sorted channel gain based CP is first performed. At each transmitting node, the power is uniformly allocated on each subcarrier. In addition, for total power constraint, the total power is also uniformly allocated to each transmitting node. Therefore, for individual power constraint, $P_{m,n} = \frac{P_{mt}}{N}$; and for total power constraint, $P_{m,n} = \frac{P_t}{MN}$; 2) *Opt. PA without CP*: only power allocation is optimized but no pairing, i.e., the same incoming and outgoing channels are assumed; 3) *Uniform PA without CP*: the same incoming and outgoing channels are assumed, then uniform PA as in the case 1 is used.

We use an OFDM system as an example of a multichannel system, and refer each subcarrier as a channel in this case. For the multihop setup, equal distance is assumed from hop to hop, and is denoted by $d_r$. No direct link between source and destination is available. The spatial reuse factor is set to $F_s = 3$ (i.e., interference is assumed negligible three hops away). We assume $M = 4$, unless it is otherwise specified. An $L$-tap frequency-selective fading channel is assumed for each hop. We define the *average SNR* as the average received SNR over each subcarrier at each receiving node under uniform power allocation. Specifically, it is defined for different power constraint as follows: under the total power constraint, $\text{SNR}_{\text{avg}} \triangleq \frac{P_t d_r^{-\alpha}}{MN\sigma^2}$, where $\alpha$ denotes the pathloss exponent and $\sigma^2$ the noise variance; under the individual power constraint, $\text{SNR}_{\text{avg}} \triangleq \frac{P_{mt} d_r^{-\alpha}}{N\sigma^2}$.

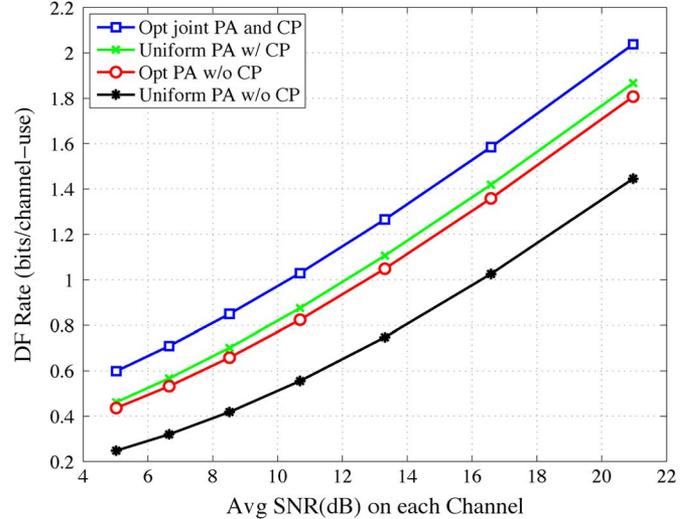

Fig. 4. Normalized rate versus the average SNR for DF OFDM relaying with $M = 4$ and $N = 64$ under total power constraint.

### A. Impact of the Average SNR

We compare the performance of various CP and PA schemes at different average SNR levels. Fig. 4 shows the normalized source-to-destination per-subcarrier rate versus the average SNR, for DF relaying under the total power constraint. The number of channels is set to $N = 64$. We observe that joint optimization of CP and PA provides significant performance improvement over the other schemes. In particular, compared with uniform PA without CP, the optimal CP alone provides 4 dB gain, and subsequently optimally allocating power provides an additional 1.5–2 dB gain. Interestingly, it is evident that channel pairing alone provides more performance gain than power allocation alone does.

Fig. 5 plots the normalized source-to-destination per-subcarrier rate versus the average SNR for AF relaying under the total power constraint. Again, $N = 64$ is used. For schemes with PA optimization, the upper-bound $R_t^{\text{up}}$ in (33) is used to obtain the PA solution. The actual rate $R_t$ obtained (as in the objective function in (32)) with such PA solution provides a lower bound on the rate under the optimal PA. In Fig. 5, for the jointly optimal CP and PA scheme and optimal PA without CP scheme, we plot both upper bound and lower bound of the rate for the optimal PA solution. We see that these two bounds become tighter as the average SNR increases, due to the improving accuracy of approximation $R_t^{\text{up}}$. The PA solution derived using $R_t^{\text{up}}$ becomes near optimal.

Comparing the performance of different CP and PA schemes shown in Fig. 5, it is seen that similarly as in the DF case, joint optimization of CP and PA provides noticeable improvement over the other schemes. The gain mainly comes from choosing CP optimally, which provides around 2 dB gain over no CP schemes. We further observe that, with optimal CP, the gap between optimal and uniform PA vanishes at higher SNR, indicating that uniform PA achieves the optimal performance at a moderately high SNR range (around 15 dB). Interestingly, this is not the case for the schemes *without CP*. The intuition behind this is the following: At relatively high SNR, it is known that the water-filling PA solution in (14) approaches a uniform allocation. Thus, the total power is approximately equally distributed to different paths. The power on each path $P_i^*$ is then further assigned optimally to each channel on the path according to (18), which is typically not uniform. The exception is when each hop exhibits a similar channel gain. This is more likely to occur as a result of channel pairing, where channels with the same rank, more likely with similar strength, are paired with each other. Therefore, with CP, the optimal PA approaches to a uniform allocation at a faster rate with increasing SNR[4]. This interesting observation suggests that, because of CP, at moderately high SNR, we are able to reduce the centralized PA solution to a simple uniform PA which requires no global channel information without losing much optimality. Note that the same argument is applicable to DF relaying, but the optimal PA approaches to a uniform allocation at a much slower rate than that for AF, which can be shown by comparing (16) and (18). The range of SNR values under consideration is too small to see the same effect in Fig. 4.

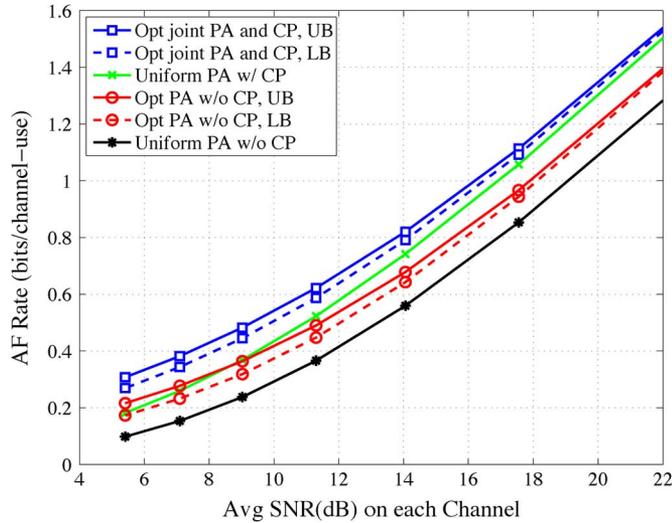

Fig. 5. Normalized rate versus the average SNR for AF OFDM relaying with $M = 4$ and $N = 64$ under total power constraint.

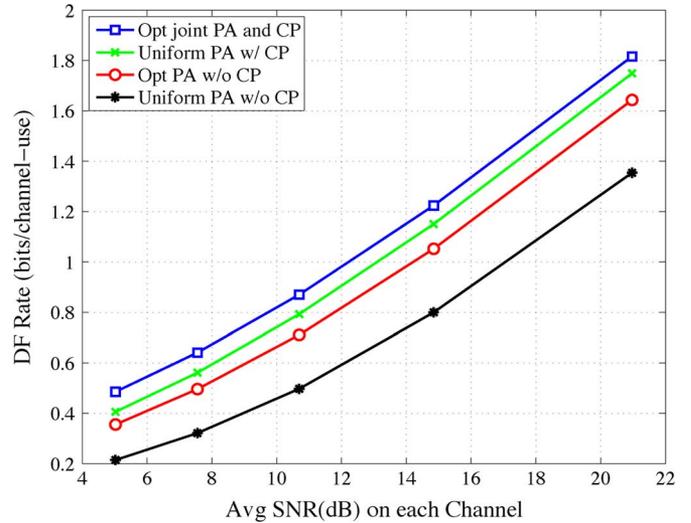

Fig. 6. Normalized rate versus the average SNR for DF OFDM relaying with $M = 4$ and $N = 16$ under individual power constraint.

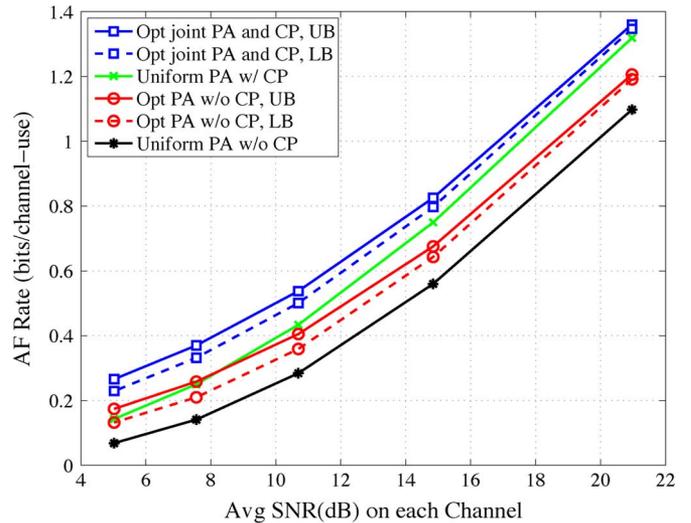

Fig. 7. Normalized rate versus the average SNR for AF OFDM relaying with $M = 4$ and $N = 16$ under individual power constraint.

Under individual power constraints, the performance comparison of CP and PA schemes are given in Figs. 6 and 7 for DF and AF relaying, respectively. We assume $N = 16$. These figures further demonstrate the significant improvement by jointly optimizing CP and PA, where most of the gain comes from optimal CP. In addition, under AF relaying, we again observe a near-optimal performance by uniform PA with CP at moderately high SNR. This suggests that, under individual power constraints, the optimal PA is close to a uniform allocation at high SNR as well when CP is adopted. This potentially simplifies greatly the PA implementation to achieve the optimal performance.

### B. Impact of the Variation of Channel Gain

In this experiment, we show how the level of channel gain variation across $N$ channels affects the performance of various CP and PA schemes. Towards this goal, we increase the number of taps of the time-domain frequency-selective channel (i.e., the maximum delay of the frequency-selective channel). This increases the level of variation of the corresponding frequency response. Figs. 8 and 9 plot the normalized per-subcarrier rate versus the number of taps of the frequency-selective channel for DF and AF relaying, respectively. The number of subcarriers is set to $N = 64$ and $\mathrm{SNR_{avg}} = 12$ dB. As we see, the performance gap between the schemes with optimal CP and without CP increases as the level of channel gain variation increases. This demonstrates that the optimal CP schemes benefit from an increased level of channel diversity, which is utilized effectively through the channel pairing. On the other hand, the relative gain of optimal PA to uniform PA is insensitive to such change and remains constant.

### C. Impact of the Number of Channels

In this experiment, we examine the effect of the number of channels, under the same level of channel gain variation across channels, on the performance of various CP and PA schemes.

---

[4]Note that, for water-filling PA, as SNR $\rightarrow \infty$, it approaches to a uniform allocation in all schemes with or without CP. The difference is the rate at which PA approaches to a uniform allocation.

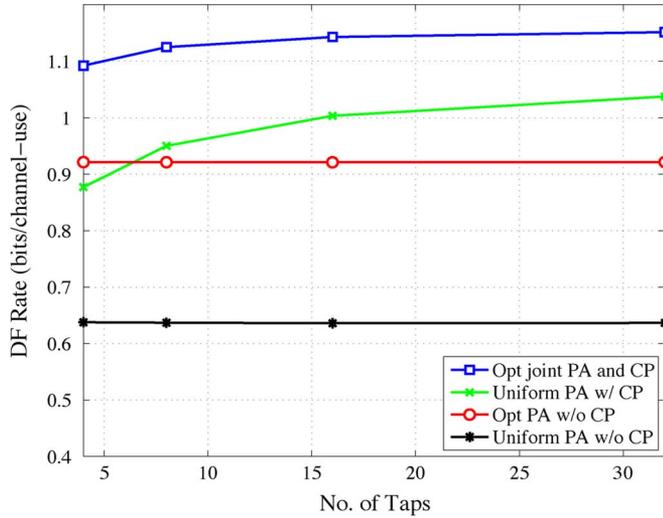

Fig. 8. Normalized rate versus number of taps for DF OFDM relaying with $M = 4$, $N = 64$, and $\mathrm{SNR_{avg}} = 12$ dB under total power constraint.

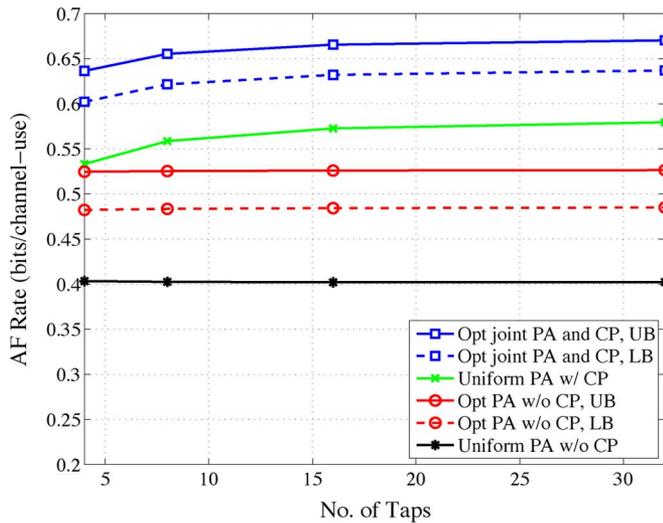

Fig. 9. Normalized rate versus number of taps for AF OFDM relaying with $M = 4$, $N = 64$, and $\mathrm{SNR_{avg}} = 12$ dB under total power constraint.

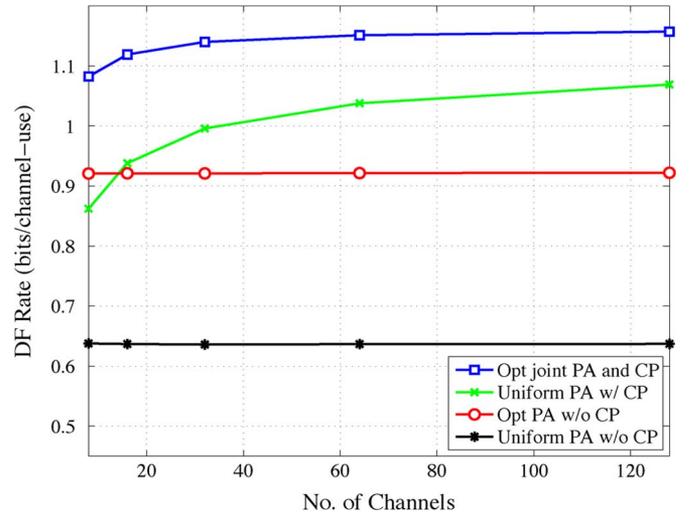

Fig. 10. Normalized rate versus number of channels for DF OFDM relaying with $M = 4$ and $\mathrm{SNR_{avg}} = 12$ dB under total power constraint.

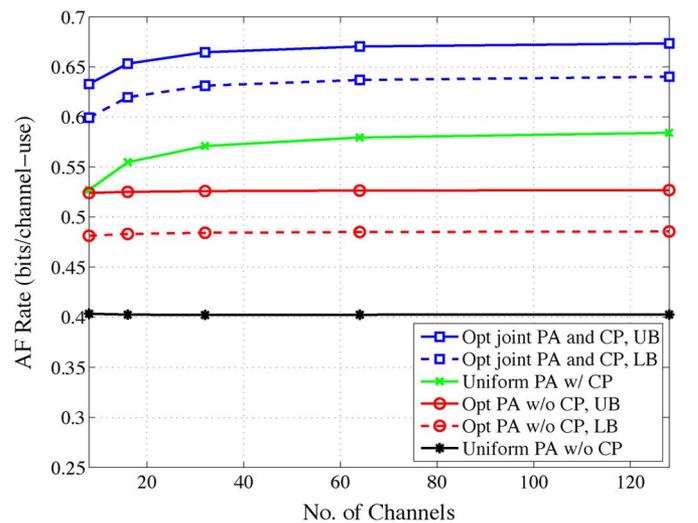

Fig. 11. Normalized rate versus number of channels for AF OFDM relaying with $M = 4$ and $\mathrm{SNR_{avg}} = 12$ dB under total power constraint.

For different $N$, the subcarrier spacing (i.e., bandwidth of each channel) is fixed. In order to set the same level of channel gain variation in frequency, we keep the maximum delay of the time-domain frequency-selective channel unchanged. Figs. 10 and 11 demonstrate the normalized per-subcarrier rate with respect to $N$ for DF and AF relaying, respectively. The average SNR is set to $\mathrm{SNR_{avg}} = 12$ dB. We observes that the gap between the two sets of schemes, with and without CP, widens as the number of channels increases. The reason behind this observation is that, as more channels becomes available, they can be exploited more judiciously for pairing, and, therefore, more gain is achieved by CP. The different PA schemes are not sensitive to the change of $N$.

### D. Impact of the Number of Hops

In this experiment, we study how the number of hops affects the performance of various CP and PA schemes. For this purpose, we increase the number of hops while keeping the distance between each hop unchanged. Figs. 12 and 13 illustrate the normalized per-subcarrier rate versus the number of hops with total power constraint for DF and AF relaying, respectively. We set $N = 64$ and $\mathrm{SNR_{avg}} = 12$ dB. As expected, for all schemes, the normalized per-subcarrier rate decreases as the number of hops increases. For DF, this is because on average the minimum rate among all hops decreases as the number of hops increases; for AF, the rate decreases due to noise amplification over hops. Comparing different schemes, we observe that the performance of the jointly optimized CP and PA scheme has the slowest decay rate, and the performance of the schemes with CP decay is slower than those without CP. In other words, the gain of optimal CP and PA is more pronounced as the number of hops increases. A multiple-fold gain is observed at a higher number of hops.

## VIII. CONCLUSION

In this paper, we have studied the problem of jointly optimizing spectrum and power allocation to maximize the source-to-destination sum rate for a multichannel $M$-hop relaying network. For fixed power allocation, we have shown that the general CP problem over multiple hops can be decomposed into $(M - 1)$ independent CP problems at each relay, where the sorted-SNR CP scheme is optimal. We then proved that a jointly optimal solution for the CP and PA problems can be achieved by decomposing the original problem into two separate CP and PA problems solved independently. It follows that the CP problem in the joint optimization can be again decomposed into independent CP problems at each relay. The solution obtained through the separate optimization bears considerably lower computational complexity compared with exhaustive-method alternatives. The separation principle is shown to hold for a variety of scenarios including AF and DF relaying strategies under either total or individual power constraints. For all these scenarios, the optimal CP scheme maps the incoming and outgoing channels at each relay according to their channel gain order, independent of the optimal PA solution. The solution for PA optimization under the individual power constraints is derived for both AF and DF relaying. Finally, we show that the above results can be directly extended to the multidestination scenario for sum-rate maximization. Significant gains in data rate were demonstrated by employing jointly optimal CP and PA in multichannel multihop relaying. It was also observed that more gain is obtained from optimal CP than optimal PA through judiciously exploiting variation among multiple channels.

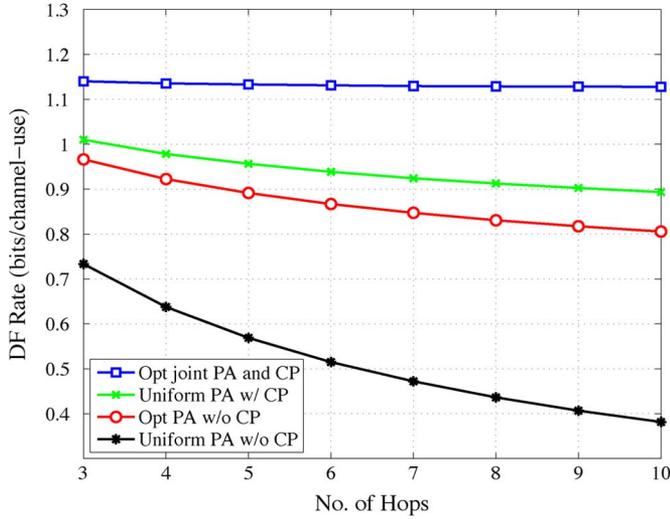

Fig. 12. Normalized rate versus number of hops for DF OFDM relaying with $N = 64$ and $\mathrm{SNR_{avg}} = 12$ dB under total power constraint.

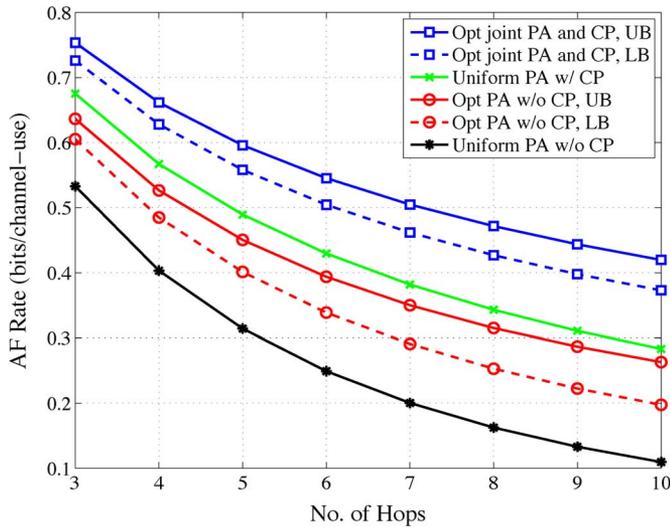

Fig. 13. Normalized rate versus number of hops for AF OFDM relaying with $N = 64$ and $\mathrm{SNR_{avg}} = 12$ dB under total power constraint.

## Appendix I
## Proof of Proposition 1

At relay 1, there are two ways to pair the channels: (1) channels 1 and 2 over hop 1 are matched with channels 1 and 2 over hop 2, respectively; (2) channels 1 and 2 over hop 1 are matched with channels 2 and 1 over hop 2, respectively. These two ways of pairing lead to the following two sets of disjoint paths from the source to the destination: $\{\mathcal{P}_i^{(1)}\} = \{(1, 1, c(3, 1)), (2, 2, c(3, 2))\}$ and $\{\mathcal{P}_i^{(2)}\} = \{(1, 2, c(3, 1)), (2, 1, c(3, 2))\}$, where the superscript $j$ in $\{\mathcal{P}_i^{(j)}\}$ indicates a different set of path selection.

By considering the *equivalent* channels from the source to the second relay, using the existing optimality result for dual-hop relaying [8], it is easy to see that $c(3, 1) = 1$ and $c(3, 2) = 2$ are optimal for $\{\mathcal{P}_i^{(1)}\}$. Furthermore, we only need to show

$$\log_2\left(1 + \gamma_{\mathrm{SD}}\left(\mathcal{P}_1^{(1)}\right)\right) + \log_2\left(1 + \gamma_{\mathrm{SD}}\left(\mathcal{P}_2^{(1)}\right)\right)$$
$$\geq \log_2\left(1 + \gamma_{\mathrm{SD}}\left(\mathcal{P}_1^{(2)}\right)\right) + \log_2\left(1 + \gamma_{\mathrm{SD}}\left(\mathcal{P}_2^{(2)}\right)\right) \quad (34)$$

for the case of $c(3, 1) = 1$ and $c(3, 2) = 2$ for both $\{\mathcal{P}_i^{(1)}\}$ and $\{\mathcal{P}_i^{(2)}\}$, since the case of $c(3, 1) = 2$ and $c(3, 2) = 1$ for $\{\mathcal{P}_i^{(2)}\}$ can be similarly proven. Inequality (34) for the AF and DF relaying cases are separately proven as follows:

*a) AF Relaying:* By inserting (1) into inequality (34) we need to show

$$\left(1 + \left(Q_1^{(1)} - 1\right)^{-1}\right)\left(1 + \left(Q_2^{(1)} - 1\right)^{-1}\right)$$
$$\geq \left(1 + \left(Q_1^{(2)} - 1\right)^{-1}\right)\left(1 + \left(Q_2^{(2)} - 1\right)^{-1}\right) \quad (35)$$

where

$$Q_1^{(1)} = \left(1 + \frac{1}{\gamma_{1,1}}\right)\left(1 + \frac{1}{\gamma_{2,1}}\right)\left(1 + \frac{1}{\gamma_{3,1}}\right)$$
$$Q_2^{(1)} = \left(1 + \frac{1}{\gamma_{1,2}}\right)\left(1 + \frac{1}{\gamma_{2,2}}\right)\left(1 + \frac{1}{\gamma_{3,2}}\right),$$
$$Q_1^{(2)} = \left(1 + \frac{1}{\gamma_{1,1}}\right)\left(1 + \frac{1}{\gamma_{2,2}}\right)\left(1 + \frac{1}{\gamma_{3,1}}\right)$$
$$Q_2^{(2)} = \left(1 + \frac{1}{\gamma_{1,2}}\right)\left(1 + \frac{1}{\gamma_{2,1}}\right)\left(1 + \frac{1}{\gamma_{3,2}}\right). \quad (36)$$

The following lemma is used to prove (35)

*Lemma 2:* With (10), we have

$$\left(Q_1^{(1)} - 1\right)\left(Q_2^{(1)} - 1\right) \leq \left(Q_1^{(2)} - 1\right)\left(Q_2^{(2)} - 1\right). \quad (37)$$

*Proof:* By substituting (36) in the following term and expanding it, we have

$$
\left(Q_1^{(1)}-1\right)\left(Q_2^{(1)}-1\right)-\left(Q_1^{(2)}-1\right)\left(Q_2^{(2)}-1\right)
$$

$$
= Q_1^{(2)}+Q_2^{(2)}-Q_1^{(1)}-Q_2^{(1)} \tag{38}
$$

$$
= \left(1+\frac{1}{\gamma_{1,1}}\right)\left(1+\frac{1}{\gamma_{2,2}}\right)\left(1+\frac{1}{\gamma_{3,1}}\right)
$$

$$
+ \left(1+\frac{1}{\gamma_{1,2}}\right)\left(1+\frac{1}{\gamma_{2,1}}\right)\left(1+\frac{1}{\gamma_{3,2}}\right)
$$

$$
- \left(1+\frac{1}{\gamma_{1,1}}\right)\left(1+\frac{1}{\gamma_{2,1}}\right)\left(1+\frac{1}{\gamma_{3,1}}\right)
$$

$$
- \left(1+\frac{1}{\gamma_{1,2}}\right)\left(1+\frac{1}{\gamma_{2,2}}\right)\left(1+\frac{1}{\gamma_{3,2}}\right)
$$

$$
= \left(\frac{1}{\gamma_{2,1}}-\frac{1}{\gamma_{2,2}}\right)
$$

$$
\times \left[\left(1+\frac{1}{\gamma_{3,2}}\right)\left(1+\frac{1}{\gamma_{1,2}}\right)\right.
$$

$$
\left.- \left(1+\frac{1}{\gamma_{3,1}}\right)\left(1+\frac{1}{\gamma_{1,1}}\right)\right]
$$

$$
\le 0 \tag{39}
$$

where we have used the fact that $Q_1^{(1)}Q_2^{(1)}=Q_1^{(2)}Q_2^{(2)}$ to arrive at (38). From (10), the first product term in (39) is negative and the second product term is positive, and therefore we obtain the last inequality. ∎

Consider the subtraction of the RHS from the LHS of (35) [see (40)–(41) at the bottom of the page], where the (40) holds because of Lemma 2, and the fact that $Q_i^{(j)}-1>0$, for $i=1,2$ and $j=1,2,3$; and the (41) holds because of (10).

*DF Relaying:* Inserting (3) into inequality (34), we need to show

$$
(1+\min(\gamma_{1,1},\gamma_{2,1},\gamma_{3,1}))(1+\min(\gamma_{1,2},\gamma_{2,2},\gamma_{3,2}))
$$
$$
\ge (1+\min(\gamma_{1,1},\gamma_{2,2},\gamma_{3,1}))(1+\min(\gamma_{1,2},\gamma_{2,1},\gamma_{3,2})). \tag{42}
$$

We can verify (42) by enumerating all possible relations among $\gamma_{m,n}$, for all $m=1,2,3$ and $n=1,2$, subject to condition (10). For example, when $\gamma_{1,1}\le\gamma_{2,1}\le\gamma_{3,1}$, $\gamma_{1,2}\le\gamma_{2,2}\le\gamma_{3,2}$, $\gamma_{2,2}\le\gamma_{1,1}\le\gamma_{3,2}$, and $\gamma_{3,2}\le\gamma_{2,1}$, (42) reduces to

$$
(1+\gamma_{1,1})(1+\gamma_{1,2}) \ge (1+\gamma_{2,2})(1+\gamma_{1,2}).
$$

The above inequality clearly holds based on the assumption of $\{\gamma_{i,j}\}$ relations. Inequality (42) can be similarly verified for all other $\{\gamma_{i,j}\}$ relations. The details are omitted for brevity.

## APPENDIX II
## PROOF OF LEMMA 1

Let $R_n$ denote the end-to-end data rate on path $n$, we have $R_n=\frac{1}{F_s}\log_2(1+\gamma_n)$, $n=1,\dots,N$, where

$$
\gamma_n = \left(\sum_{m=1}^{M}\frac{1}{P_{m,n}a_{m,n}}\right)^{-1}. \tag{43}
$$

Then $R_t^{\mathrm{up}}=\sum_{n=1}^{N}R_n$. To show $R_t^{\mathrm{up}}$ is concave in $\{P_{m,n}\}$, it suffices to show that each $R_n$ is concave in $\{P_{mn}\}$. The concavity proof of $R_n$ follows the concavity of $\gamma_n$ due to the composition rules which preserve concavity [23]. For simplicity, we drop the subscript $n$ from notations in (43). In the following we prove that $\gamma(\bar{P})$ is concave in $\bar{P}$, where $\bar{P}=[P_1,\dots,P_M]^T$. The second-order partial derivatives of $\gamma(\bar{P})$ are given by

$$
\frac{\partial^2\gamma(\bar{P})}{\partial P_j^2} = \frac{-2}{a_jP_j^3}\left(\sum_{i=1}^{M}\frac{1}{a_iP_i}\right)^{-2}
$$
$$
+ \frac{2}{a_j^2P_j^4}\left(\sum_{i=1}^{M}\frac{1}{a_iP_i}\right)^{-3} \tag{44}
$$
$$
\text{and}
$$
$$
\frac{\partial^2\gamma(\bar{P})}{\partial P_j\partial P_k} = \frac{1}{a_jP_j^2}\frac{2}{a_kP_k^2}\left(\sum_{i=1}^{M}\frac{1}{a_iP_i}\right)^{-3}, \quad \text{for } k\ne j. \tag{45}
$$

---

LHS of (35) - RHS of (35)

$$
= \underbrace{\left(\left(Q_1^{(1)}-1\right)^{-1}+\left(Q_2^{(1)}-1\right)^{-1}+\left(Q_1^{(1)}-1\right)^{-1}\left(Q_2^{(1)}-1\right)^{-1}\right)}_{A}
$$

$$
- \underbrace{\left(\left(Q_1^{(2)}-1\right)^{-1}+\left(Q_2^{(2)}-1\right)^{-1}+\left(Q_1^{(2)}-1\right)^{-1}\left(Q_2^{(2)}-1\right)^{-1}\right)}_{B}
$$

$$
\ge A\left(Q_1^{(1)}-1\right)\left(Q_2^{(1)}-1\right)-B\left(Q_1^{(2)}-1\right)\left(Q_2^{(2)}-1\right) \tag{40}
$$

$$
= Q_2^{(1)}+Q_1^{(1)}-Q_1^{(2)}-Q_2^{(2)}
$$

$$
= \left(\frac{1}{\gamma_{2,2}}-\frac{1}{\gamma_{2,1}}\right)\left[\left(1+\frac{1}{\gamma_{1,2}}\right)\left(1+\frac{1}{\gamma_{3,2}}\right)-\left(1+\frac{1}{\gamma_{1,1}}\right)\left(1+\frac{1}{\gamma_{3,1}}\right)\right]
$$

$$
\ge 0 \tag{41}
$$

Hence, the Hessian matrix $\Delta^2\gamma(\bar{P})$ can be expressed as

$$\Delta^2\gamma(\bar{P}) = \left(\sum_{i=1}^{M}\frac{1}{a_iP_i}\right)^{-3}\left(-2\left(\sum_{i=1}^{M}\frac{1}{a_iP_i}\right)\right.$$
$$\left.\times \operatorname{diag}\left(\frac{1}{a_1P_1^3},\ldots,\frac{1}{a_MP_M^3}\right) + 2\mathbf{q}\mathbf{q}^T\right) \quad (46)$$

where $\mathbf{q} = [q_1,\ldots,q_M]^T$ with $q_m = \frac{1}{a_mP_m^2}, m = 1,\ldots,M$, and $\operatorname{diag}(\mathbf{x})$ denotes the diagonal matrix with diagonal elements being the elements in vector $\mathbf{x}$. To prove concavity, we need to show $\Delta^2\gamma(\bar{P}) \preceq 0$. For any vector $\mathbf{v} = [v_1,\ldots,v_M]^T$, we have

$$\mathbf{v}^T\Delta^2\gamma(\bar{P})\mathbf{v} = 2\left(\sum_{i=1}^{M}\frac{1}{a_iP_i}\right)^{-3}$$
$$\times\left(-\left(\sum_{i=1}^{M}\frac{1}{a_iP_i}\right)\sum_{i=1}^{M}\frac{1}{a_iP_i^3}v_i^2\right.$$
$$\left.+\left(\sum_{i=1}^{M}\frac{v_i}{a_iP_i^2}\right)^2\right)$$
$$\leq 0$$

where the inequality is obtained by using the Cauchy-Schwartz inequality $(\mathbf{e}^T\mathbf{e})(\mathbf{c}^T\mathbf{c}) \geq (\mathbf{e}^T\mathbf{c})^2$ for two vectors $\mathbf{e} = [e_1,\ldots,e_M]^T$ and $\mathbf{c} = [c_1,\ldots,c_M]^T$, with $e_i = \frac{1}{\sqrt{a_iP_i}}$ and $c_i = \frac{v_i}{\sqrt{a_iP_i^3}}$. Therefore, $\Delta^2\gamma(\bar{P}) \preceq 0$.

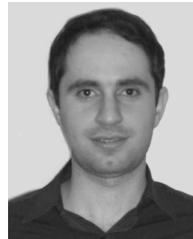

**Mahdi Hajiaghayi** (S'10) received the B.Sc. degrees in electrical engineering and petroleum engineering simultaneously from Sharif University of Technology, Tehran, Iran, in 2005, and the M.Sc. degree in electrical and computer engineering from the University of Alberta, Edmonton, Alberta, Canada, in 2007.

He is currently pursuing the Ph.D. degree in electrical and computer engineering at the University of Toronto, Toronto, Ontario, Canada. His research interests include resource allocation in wireless cooperative and sensor networks, and investigating the application of optimization techniques in wireless communication.

Mr. Hajiaghayi received the Ted Rogers scholarship and is a member of the IEEE communication society. He also won the 2010–2011 Ontario Graduate Scholarship (OGS).

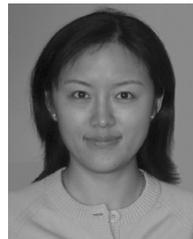

**Min Dong** (S'00–M'05–SM'09) received the B.Eng. degree from Tsinghua University, Beijing, China, in 1998, and the Ph.D. degree in electrical and computer engineering with minor in applied mathematics from Cornell University, Ithaca, NY, in 2004.

During 2004–2008, she was with Corporate Research and Development, Qualcomm Inc., San Diego, CA, where she has actively contributed to the research, design, and standardization of current and future OFDMA technologies for the 4th-generation broadband mobile communications. Since July 2008, she has been with the Faculty of Engineering and Applied Science, University of Ontario Institute of Technology, Ontario, Canada, where she is currently an Assistant Professor. She also holds a status-only Assistant Professor appointment with the Department of Electrical and Computer Engineering, University of Toronto. Her research interests are in the areas of adaptive signal processing for communications, cooperative and adaptive communication and networking techniques and optimization, and broadband mobile access networks.

Dr. Dong received the 2004 Best Paper Award from IEEE Signal Processing Society. She currently serves as an Associate Editor for the IEEE TRANSACTIONS ON SIGNAL PROCESSING and IEEE SIGNAL PROCESSING LETTERS.

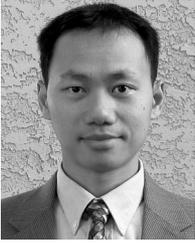

**Ben Liang** (S'94–M'01–SM'06) received the B.Sc. and M.Sc. degrees with simultaneous honors (valedictorian) in electrical engineering from Polytechnic University in Brooklyn, New York, in 1997, and the Ph.D. degree in electrical engineering with a computer science minor from Cornell University, Ithaca, NY, in 2001.

During the 2001–2002 academic year, he was a Visiting Lecturer and Postdoctoral Research Associate with Cornell University. He joined the Department of Electrical and Computer Engineering, University of Toronto, Toronto, Canada, in 2002, where he is now an Associate Professor. His current research interests are in mobile networking and multimedia systems.

Dr. Liang received an Intel Foundation Graduate Fellowship in 2000 and an Ontario MRI Early Researcher Award (ERA) in 2007. He was a coauthor of the Best Paper Award at the IFIP Networking Conference in 2005. He is an editor for the IEEE TRANSACTIONS ON WIRELESS COMMUNICATIONS and an Associate Editor for the Wiley *Security and Communication Networks* journal. He serves on the organizational and technical committees of a number of conferences each year. He is a member of ACM and Tau Beta Pi.